\documentclass[%
 reprint,
superscriptaddress,
longbibliography,
 amsmath,amssymb,
 aip,
 citeautoscript,
]{revtex4-1}

\usepackage{graphicx}
\usepackage{dcolumn}
\usepackage{bm}
\usepackage{amsmath}
\usepackage{amssymb}
\usepackage{amsthm}
\usepackage{units}
\usepackage{gensymb}
\usepackage{xcolor}
\newcommand{\tN}{t_\mathrm{N}}
\newcommand{\kB}{k_\mathrm{B}}

\begin{document}

\title{Distinct chemistries explain decoupling of slip and wettability in atomically smooth aqueous interfaces}

\author{Anthony R. Poggioli}
\affiliation{Department of Chemistry, University of California, Berkeley}
\affiliation{Kavli Energy NanoScience Institute, Berkeley, California}

\author{David T. Limmer}
\email{dlimmer@berkeley.edu}
\affiliation{Department of Chemistry, University of California, Berkeley}
\affiliation{Kavli Energy NanoScience Institute, Berkeley, California}
\affiliation{Materials Science Division, Lawrence Berkeley National Laboratory}
\affiliation{Chemical Science Division, Lawrence Berkeley National Laboratory}

\date{\today}

\begin{abstract}
Despite essentially identical crystallography and equilibrium structuring of water, nanoscopic channels composed of hexagonal boron nitride and graphite exhibit an order-of-magnitude difference in fluid slip. We investigate this difference using molecular dynamics simulations, demonstrating that its origin is in the distinct chemistries of the two materials. In particular, the presence of polar bonds in hexagonal boron nitride, absent in graphite, leads to Coulombic interactions between the polar water molecules and the wall. We demonstrate that this interaction is manifested in a large typical lateral force experienced by a layer of oriented hydrogen atoms in the vicinity of the wall, leading to the enhanced friction in hexagonal boron nitride. The fluid adhesion to the wall is dominated by dispersive forces in both materials, leading to similar wettabilities. Our results rationalize recent observations that the difference in frictional characteristics of graphite and hexagonal boron nitride cannot be explained on the basis of the minor differences in their wettabilities.
\end{abstract}
\maketitle

Nanofluidics is the basis for emerging membrane-based desalination and blue energy generation technologies \cite{Bocquet_Charlaix2010, coming_of_age, Picallo_et_al2013, Yang_et_al2018, Zhang_et_al2015, Laucirica_et_al2020,Siria_et_al2017}. The viability of such technologies is limited by the attainable efficiency of flow generation and power conversion \cite{Siria_et_al2017}. The frictional dissipation, characterized by the degree to which fluid slips past a solid-liquid interface, thus enters as a fundamental parameter in need of molecular characterization \cite{Huang_slip}. Two paradigmatic materials in this context are hexagonal boron nitride (HBN) and layered graphite (GR), which, while crystallographically identical, exhibit nearly an order-of-magnitude difference in slip, independent of macroscopic geometry\cite{Radha2018_hydrodynamic, Kuman_Kannam2012, Mouterde_et_al2019, Radha2016, Tocci_et_al2014, Secchi_et_al2016_nature, Tocci_et_al2020, Secchi_et_al2016_nature,Radha2016}. In order to rationalize this difference, we have employed molecular dynamics simulations of channels composed of HBN and GR, finding that the origin of the difference is in the distinct chemistries of the two materials. The polar bonds of HBN, absent in GR, lead to a nearly order-of-magnitude increase in the variance of the lateral force experienced by the water from the wall. This difference is not manifested in the vertical component of the electrostatic wall force, leading to a dominance of dispersion forces in the attractive water-wall interaction, and hence to similar wettabilities of the two materials. This asymmetry explains the decoupling of wettability properties from frictional characteristics in HBN and other heteroatomic materials containing polar bonds \cite{Tocci_et_al2014, Tocci_et_al2020, Rajan_hBN_electrostatics}.

The no-slip condition, ubiquitous in macroscopic fluid mechanics \cite{Bocquet_Charlaix2010, Kirby2010, Landau_Lifshitz_FM}, posits that the tangential velocity in the vicinity of a wall exactly vanishes. However, in nanofluidics, one must allow for the finite slip of fluid past a solid boundary, reflecting a deviation of nanoscale fluid transport from classical macroscopic hydrodynamic theory due to the predominance of interfacial effects. This finite slip is characterized by the slip length $b$ \cite{Bocquet_Barrat1994, Bocquet_Barrat2007_review}, which relates the average fluid velocity $\langle u(z) \rangle$ in the vicinity of a wall to the velocity gradient normal to the wall $\partial_z \langle u \rangle |_{\rm wall}$ via the partial-slip boundary condition
\begin{equation}
\left.\langle u \rangle\right|_{\rm wall} = b\, \partial_z \langle u \rangle |_{\rm wall},
\label{eqn:slip_defn}
\end{equation}
where $\left.\langle u \rangle\right|_{\rm wall}$ is the fluid velocity at the wall. The slip length is a coefficient of proportionality with the unit of length and may be interpreted geometrically as the distance beyond the wall at which the fluid velocity extrapolates to zero \cite{Kirby2010}.

Understanding the molecular origins of slip in nanometric confinement has captured consistent attention over the last few decades \cite{Bocquet_Barrat1994, Barrat_Bocquet1999_scaling, Charlaix_2005_exp, Kuman_Kannam2012, Alvarado_et_al2016, Xu_et_al2018}, yet there is no satisfactory theoretical prediction for the slip length from fundamental microscopic parameters of the fluid and material. Much theoretical, numerical, and experimental work has focused on the role of wettability as quantified by the contact angle in fluid slip \cite{Bakli_and_Chakraborty_2019, Charlaix_2005_exp, Huang_Bocquet2008_scaling, Xu_et_al2018, Bocquet_Barrat1994, Barrat_Bocquet1999_scaling}. However, recent work has indicated that the differences in wettability between GR and HBN are relatively minor and cannot explain the dramatic difference in solid-liquid frictions \cite{Tocci_et_al2014, Tocci_et_al2020, Rajan_hBN_electrostatics, Wei_and_Luo_2018, Li_and_Zeng_2012, wetting_ref3}. In addition to this fundamental theoretical challenge, the presence of large slip lengths in certain materials, for example graphene and carbon nanotubes \cite{Secchi_et_al2016_nature,Radha2016,Mouterde_et_al2019}, presents the promise of designing high efficiency membranes for desalination and blue energy generation. A greater degree of fluid slip translates to diminished frictional energy dissipation and hence to greatly enhanced flow generation or power conversion for a given input of electrical or osmotic energy \cite{Huang_slip,Siria_et_al2017,coming_of_age}.

\begin{figure}[t]
	\centerline{\includegraphics[scale=0.29]{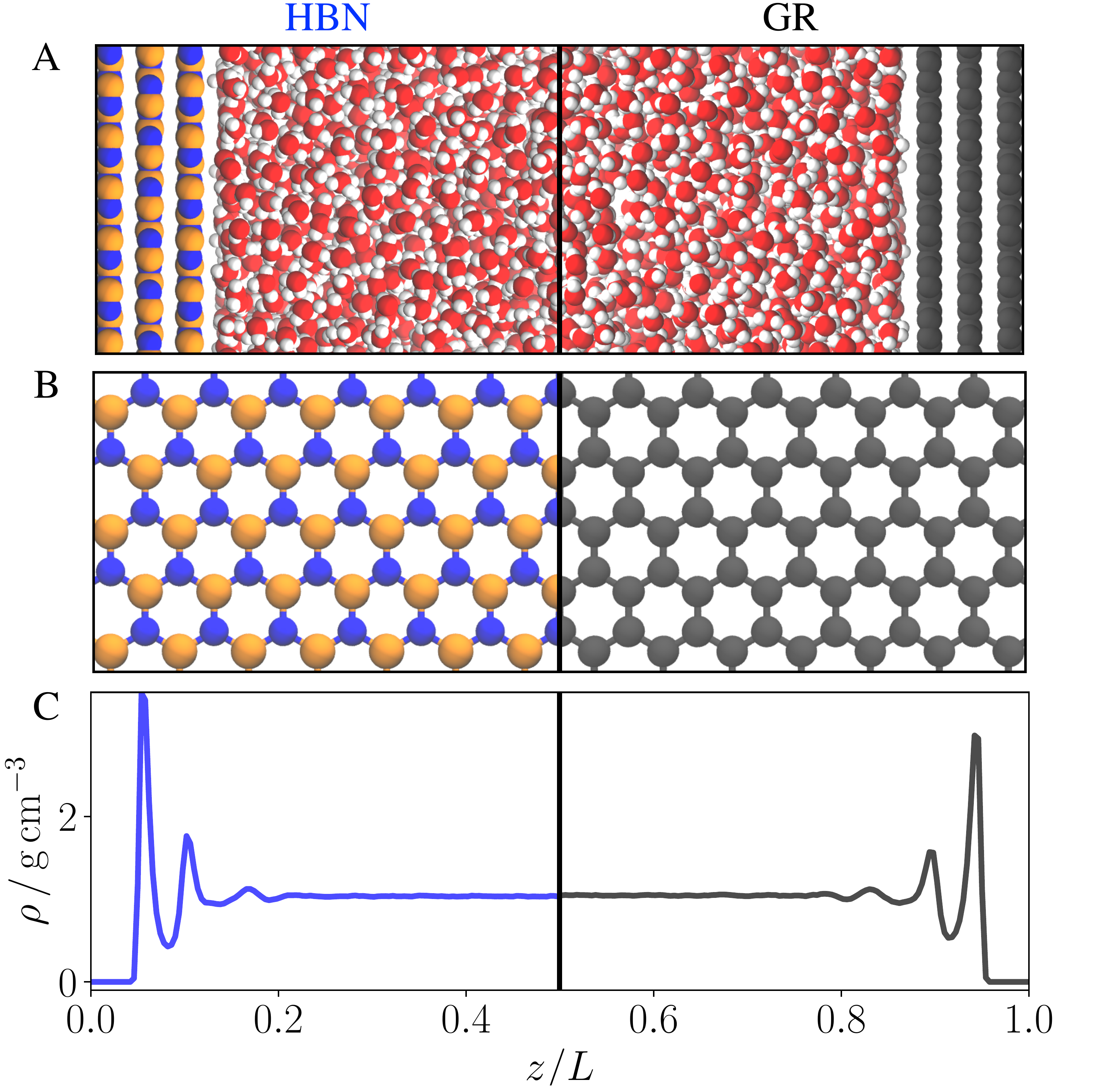}}
    \caption{Model configuration and equilibrium water density. A) For both HBN (left) and GR (right), we consider channels composed of two walls separated by approximately $6 \, {\rm nm}$ and composed of three layers. The channels are filled with TIP4P/2005 water. Red, white, orange, blue, and black spheres represent, respectively, oxygen, hydrogen, boron, nitrogen, and carbon atoms. B) Lattices of single HBN and GR layers. C) The equilibrium density of water confined in HBN (blue) and GR (black). Blue and black curves will indicate HBN and GR, respectively, throughout.}
	\label{fig:setup}
\end{figure}

In order to study slip in HBN and GR, we use molecular dynamics simulations. Figure \ref{fig:setup}A shows snapshots of our systems. We consider HBN and GR channels of $L \approx 6 \, \rm nm$ in width, defined as the distance between the center-of-mass location of the interior wall layers. The channels are bounded by two fixed walls of area $A \approx 3 \times 3 \, {\rm nm}^2$, each composed of three atomic layers. The channel widths are adjusted such that the bulk density $\bar{\rho} \approx 1  \, {\rm g} / {\rm cm}^{3}$. Periodic boundary conditions are imposed in all three directions, with a vacuum layer of roughly $1.5 \, {\rm nm}$ beyond the walls to ensure that the periodic images of the channel do not interact with one another. The channels are filled with $N = 2000$ TIP4P/2005 water molecules \cite{TIP4P2005} with rigid geometries imposed by the SHAKE algorithm \cite{SHAKE}. Lennard-Jones parameters for the $sp^2$ carbon in GR are taken from the Amber96 force field \cite{AMBER96_C}, and those for boron and nitrogen in the HBN simulations are taken from Ref. \onlinecite{Kayal_Chandra2019}. The Lennard-Jones parameters for carbon, boron, and nitrogen are very similar, reflecting the similar corrugations in the short-range repulsive wall interactions. Lennard-Jones parameters for heteroatomic interactions are determined using Lorentz-Berthelot mixing rules. Additionally, owing to the polar bond present in HBN, the boron and nitrogen atoms have formal charges of $\pm 1.05e$, respectively, which interact Coulombically with the polar water molecules \cite{Kayal_Chandra2019}.

Figure \ref{fig:setup}B shows a single layer of HBN and GR. Experimentally, the interatomic spacings in the two lattices are found to be essentially identical ($1.42 \, \rm \AA$ in GR and $1.45 \, \rm \AA$ in HBN) and are set to $1.42 \, {\rm \AA}$ in the simulations. \cite{hBN_crystallography, gr_crystallography_journal, crystallography_book} The interlayer spacings are set to $3.38 \, \rm \AA$, compared to experimental values of $3.39 \, \rm \AA$ for GR and $3.33 \, \rm \AA$ for HBN. \cite{hBN_crystallography, gr_crystallography_journal, crystallography_book} The HBN and GR layers are arranged according to their equilibrium stackings -- AA' stacking for HBN and AB (Bernal) stacking for GR \cite{hBN_crystallography, gr_crystallography_journal, crystallography_book} -- and are held rigid. All simulations were performed in LAMMPS in the NVE ensemble after equilibrating for approximately $5 \, \rm ns$ with a Langevin thermostat at $T = 298 \, \rm K$.\cite{plimpton1995fast} The total length of all simulations used is approximately $60 \, {\rm ns}$.

The identical crystal structures lead to similar equilibrium water structures. We illustrate this with the density profiles, shown in Fig. \ref{fig:setup}C. Both materials induce a strong layering of water in the vicinity of the solid-liquid interface,\cite{mccaffrey2017mechanism,strong2016atomistic} with a contact density that is slightly higher for HBN than GR. This similarity in out-of-plane structure suggests similar wettabilities, in contrast to the strong difference in frictional properties. The orientational structure of water is also similar in the two materials, with both showing a preferential orientation of water in the vicinity of the interface such that O-H bonds are orientated parallel to the wall. In HBN, there is an additional tendency for some of the O-H bonds to arrange themselves such that the H atoms are oriented towards the positively charged nitrogen centers, leading to a slightly enhanced hydrogen density and a net attractive electrostatic interaction near the wall. These aspects of the equilibrium water structure are also observed in {\it ab initio} simulations, indicating that these force field parameters accurately parameterize the solid-liquid interaction. \cite{Kayal_Chandra2019,Tocci_et_al2014}

To characterize the flow properties in these channels, we compute their permeability to water. We examine two related quantities indicating the ease with which fluid flow is generated by the application of a pressure gradient $\partial_x p$: the mobility $\mathcal{M}(z)$ and hydraulic conductivity $\mathcal{L}$. The mobility encodes the shape of the average velocity profile $\left\langle u(z) \right\rangle$ induced by an applied pressure gradient, while the hydraulic conductivity indicates the magnitude of the average velocity $\left\langle U \right\rangle = (1/L) \int {\rm d}z \, \left\langle u(z) \right\rangle$. The definitions and corresponding Green-Kubo relations for these two quantities are given by \cite{Rotenberg_mobility}
\begin{equation}
\mathcal{M}(z) \equiv \frac{\left\langle u(z) \right\rangle}{-\partial_x p} = \beta V \int_0^{\tN} {\rm d} t \, \left\langle u(z,t) U(0) \right\rangle, \quad 
\label{eqn:mobility}
\end{equation}
and
\begin{equation}
\mathcal{L} \equiv \frac{\left\langle U \right\rangle}{-\partial_x p} = \frac{1}{L} \int_0^L {\rm d} z \, \mathcal{M}(z) = \beta V \int_0^{\tN} {\rm d}t \, C_{UU}(t).
\label{eqn:conductivity}
\end{equation}
In these expressions, $\beta \equiv 1/\kB T$ is the inverse of the temperature times Boltzmann's constant, $V = AL$ is the channel volume, and $u(z,t) = (L/N) \sum_{i = 1}^N v_{i,x}(t) \delta [z - z_i(t)]$ is the instantaneous molecular velocity profile in $z$. The function $C_{UU}(t) \equiv \left\langle U(t) U(0) \right\rangle$ appearing in Eq. \ref{eqn:conductivity} is the average velocity autocorrelation, and the time $\tN$ appearing in Eqs. \ref{eqn:mobility} and \ref{eqn:conductivity} are taken long enough that the integrated correlation functions have plateaued, but not so long that they have begun to decay to zero. This decay at long times is a reflection of the finite lateral extent of these systems, and evaluating the Green-Kubo relation at the plateau time is the standard procedure in such scenarios \cite{Bocquet_Barrat1994,Falk_et_al2012,Rajan_hBN_electrostatics,nakano2020equilibrium}.

Care must be taken in determining the fluid volume in which to apply the above Green-Kubo relations. In particular, since we will ultimately connect these results to a hydrodynamic model of the fluid flow, we must properly partition the fluid volume into three subregions: a hydrodynamic region where the bulk viscosity applies \cite{Zhou_et_al2021_viscosity} and two contact regions in the vicinity of the walls. This partitioning amounts to determining the location of the hydrodynamic interface, the plane at which the boundary condition given in Eq. \ref{eqn:slip_defn} is applied \cite{Bocquet_Barrat1994}. An analysis by Chen {\it et al.} \cite{Chen_interface} has shown that the hydrodynamic interface coincides closely with the second density peak for fluids in the vicinity of either hydrophobic or hydrophilic walls, and we take this as our operational definition of the hydrodynamic interface. In this case, defining the distance of the hydrodynamic interface from the wall as $\Delta_{\rm hyd}\approx 6\mathrm{\AA}$, the system size appearing in Eqs. \ref{eqn:mobility} and \ref{eqn:conductivity} is properly taken to be $L_{\rm hyd} \equiv L - 2 \Delta_{\rm hyd}$, corresponding to a volume of $V_{\rm hyd} = A L_{\rm hyd}$ containing on average $\left\langle N_{\rm hyd} \right\rangle$ particles. It has recently been shown that such a decomposition is crucial to obtain the correct value of the bulk viscosity from the mobility and velocity profiles \cite{Zhou_et_al2021_viscosity}.

\begin{figure}[t]
	\centerline{\includegraphics[scale=0.35]{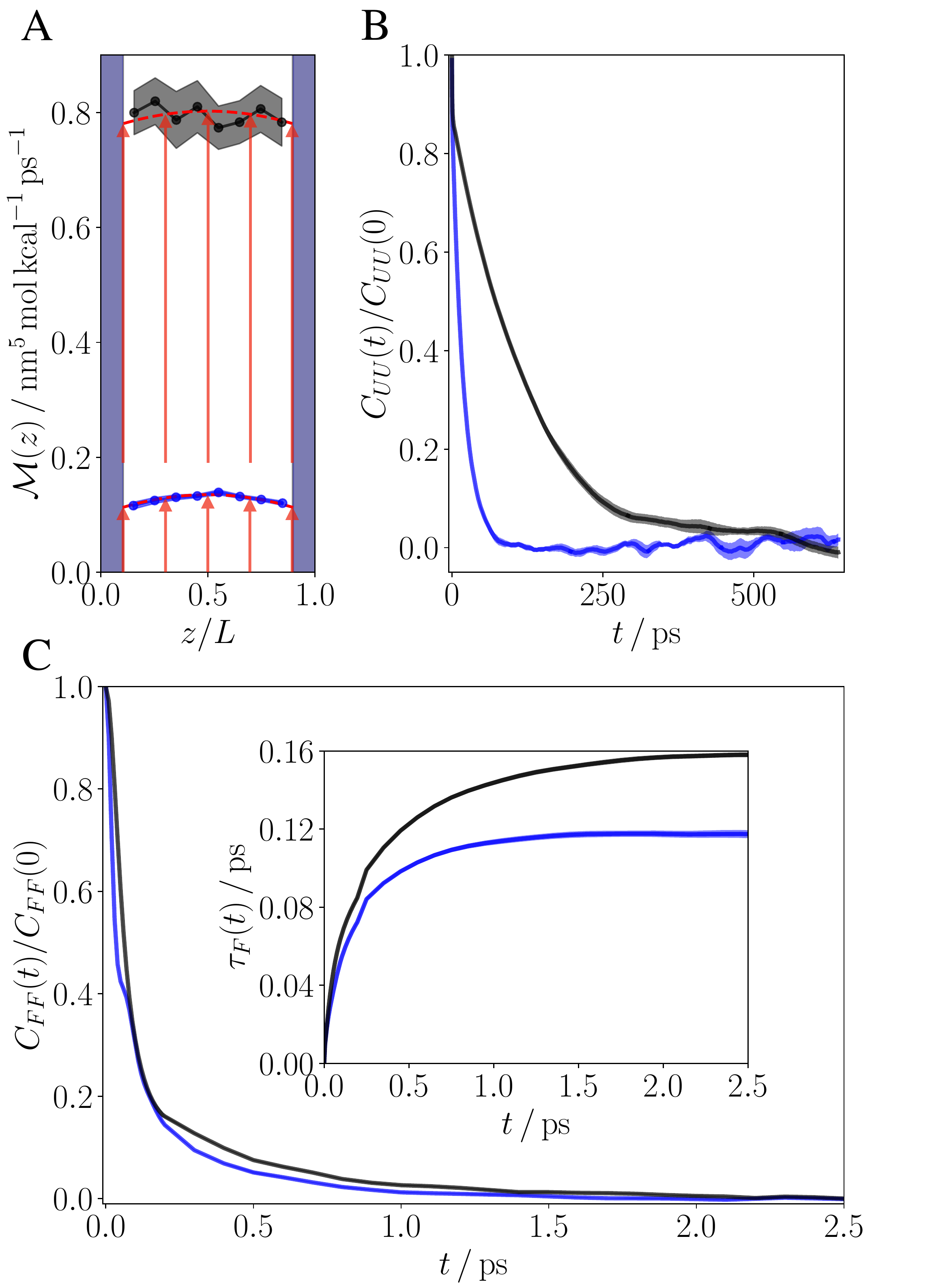}}
    \caption{Estimates of slip length in HBN (blue) and GR (black). A) The profiles of mobility $\mathcal{M}(z)$ obtained from Eq. \ref{eqn:mobility}, with fits to the Poiseuille solution (Eq. \ref{eqn:mobility_Poiseuille}) indicated by the dashed red lines. B) The average molecular velocity autocorrelations $C_{UU}(t)$. C) The wall force autocorrelations $C_{FF}(t)$, and their integrals $\tau_F(t)$ (inset). The shaded vertical regions in panel A indicate the contact regions. The error bars indicated here and in Fig. \ref{fig:contributions} are obtained by binning our data into equal-length trajectories and calculating the standard error of the means obtained for each bin.}
	\label{fig:estimates}
\end{figure}

Figure \ref{fig:estimates}A shows our estimates of the mobility profiles according to Eq. \ref{eqn:mobility}. The profiles offer a striking demonstration of the difference in flow characteristics between these otherwise quite similar materials. We see that, while HBN does exhibit some degree of finite slip past the solid boundaries, the slip in GR is much more pronounced, resulting in an essentially flat profile compared to the clearly quadratic profile observed in HBN. This is an indication that the contribution of the boundary slip to the mobility in GR greatly exceeds that of the viscous response in the bulk region.

Figure \ref{fig:estimates}B illustrates the average fluid velocity autocorrelations $C_{UU}(t)$ for both materials. Since the variances $C_{UU}(0)$ are set by equipartition and nearly identical for the two materials, the differences in the conductivities are due to the longer decay time observed for GR than for HBN. The decay of $C_{UU}(t)$ is much longer than the hydrodynamic relaxation time anticipated in the absence of interfacial slip, $\rho L_\mathrm{hyd}^2/\eta \approx 30$ ps, where $\eta$ is the bulk viscosity. This slow relaxation is a reflection of the finite slip at the solid-liquid interface, associated with an additional characteristic relaxation time scale, $\rho L_{\rm hyd} b/\eta$. Using Eq. \ref{eqn:conductivity}, we find that the conductivity in GR exceeds that in HBN by roughly a factor of six ($\mathcal{L}=0.762$ and $0.123 \, {\rm nm}^5 \, {\rm mol} \, {\rm kcal}^{-1} \, {\rm ps}^{-1}$). This difference is consistent with experimental and computational results \cite{Radha2018_hydrodynamic, Kuman_Kannam2012, Mouterde_et_al2019, Radha2016, Tocci_et_al2014, Tocci_et_al2020}.

The most straightforward comparison of our results to those obtained in experiments and other computational studies is in terms of the slip length. We can extract estimates of the slip length from our data by connecting our results to a hydrodynamic model of the pressure-induced flow -- namely, the Poiseuille solution to the Navier-Stokes equations subject to the partial-slip boundary condition in Eq. \ref{eqn:slip_defn} for the flow field generated by a uniform pressure gradient \cite{Kirby2010}. The Poiseuille solution gives for the mobility and conductivity
\begin{equation}
\mathcal{M}_P(z) = \frac{L_{\rm hyd}^2}{2\eta} \left[ \left( \frac{b}{L_{\rm hyd}} \right)^2 + \frac{z}{L_{\rm hyd}} - \left( \frac{z}{L_{\rm hyd}} \right)^2 \right], 
\label{eqn:mobility_Poiseuille}
\end{equation}
and
\begin{equation}
\mathcal{L}_P = \frac{L_{\rm hyd}^2}{12 \eta} \left( 1 + 6 \frac{b}{L_{\rm hyd}} \right).
\label{eqn:conductivity_Poiseuille}
\end{equation}
As we have an independent method of estimating $L_{\rm hyd}$ from the density profile and $\eta$ for the water model is known, the only unknown in both of these equations is the slip length $b$. We obtain two estimates of the slip from the measured value of the conductivity and a fit of the mobility profile to Eq. \ref{eqn:mobility_Poiseuille}. We give the results of these estimates in Table \ref{table:slip}. The values obtained from the two estimation methods are consistent for both materials, giving a slip length of $\sim 6 \, \rm nm$ in HBN and $\sim 40 \, \rm nm$ in GR. These values agree quantitatively with previous experimental and computational results \cite{Rajan_hBN_electrostatics,Radha2018_hydrodynamic, Kuman_Kannam2012, Mouterde_et_al2019, Radha2016, Tocci_et_al2014, Tocci_et_al2020, Bocquet_Barrat2007_review, Charlaix_2005_exp, Huang_Bocquet2008_scaling, Falk_et_al2010, Xu_et_al2018, Alvarado_et_al2016} and illustrate the approximately order-of-magnitude difference in slip characteristics between the two materials.

\begin{table}
 \begin{center}
  \begin{tabular}{c|c|c}
	slip length (nm) & HBN & GR \\ \hline
	conductivity & $5.71 \pm 0.18$ & $39.9 \pm 2.13$ \\
	mobility & $5.95 \pm 0.16$ & $41.7 \pm 2.13$	 \\
	force & $6.81 \pm 0.10$ & $40.4 \pm 0.54$ \\
  \end{tabular}
  \caption{Slip length estimated in HBN and GR using each of the three methods discussed in the text. Error bars indicate the standard error of the mean.}
  \label{table:slip}
 \end{center}
\end{table}

When fitting our mobility profiles to Eq. \ref{eqn:mobility_Poiseuille}, we assume the known bulk viscosity of TIP4P/2005 water at $298 \, \rm K$, $\eta = 0.855 \, {\rm mPa \, s}^{-1}$. This assumption is appropriate because we have partitioned the channel volume into regions dominated by the wall interaction and by the viscous fluid-fluid interaction, and the bulk viscosity holds in the latter region \cite{Zhou_et_al2021_viscosity}. The validity of this approach is confirmed by our results for the mobility in HBN (Fig. \ref{fig:estimates}A), where we see that the curvature of our fit determined by the bulk viscosity (Eq. \ref{eqn:mobility_Poiseuille}) matches the curvature measured in our simulations within the statistical uncertainty. The profile obtained in GR appears to be essentially flat because the high degree of slip in GR leads to large equilibrium fluctuations in the velocity, making it difficult to converge the mobility profile.

We may measure the slip directly, rather than extracting it as a fit parameter when comparing our data to a hydrodynamic model of the flow. To do so, we must consider the interfacial friction coefficient $\lambda$, defined by $\left\langle F_{\rm wall} \right\rangle/A = - \lambda \left.\langle u \rangle\right|_{\rm wall}$, where $F_{\rm wall}$ is the total force of the wall on the fluid. 
This parameter may be related to the slip length through a force balance $- \left\langle F_{\rm wall} \right\rangle / A = \eta \partial_z \left.\langle u \rangle\right|_{\rm wall}$. This gives $\lambda = \eta/b$, relating the slip length to the ratio of interfacial and bulk dissipation.
The Green-Kubo relationship for $\lambda$ is \cite{Bocquet_Barrat1994}
\begin{equation}
\lambda = \frac{\beta}{A} \int_0^{\tN} {\rm d}t \, C_{FF}(t) \equiv \frac{\beta }{A} \left\langle F_{\rm wall}^2 \right\rangle \tau_F(\tN),
\label{eqn:lambda}
\end{equation}
where we have introduced the wall-force autocorrelation $C_{FF}(t) \equiv \left\langle F_{\rm wall}(t) F_{\rm wall}(0) \right\rangle$. As in the Green-Kubo relations for the mobility and conductivity, it is necessary to evaluate the integral of $C_{FF}(t)$ at a plateau time $\tN$. We observe from Eq. \ref{eqn:lambda} that the interfacial friction coefficient may be decomposed into a static contribution $\left\langle F_{\rm wall}^2 \right\rangle$ and a dynamic contribution $\tau_F(\tN) \equiv \int {\rm d} t \, C_{FF}(t)/C_{FF}(0)$ \cite{Barrat_Bocquet1999_scaling, Rajan_hBN_electrostatics, Tocci_et_al2020}.

We show the results for $C_{FF}(t)$ for water on HBN and GR in Fig. \ref{fig:estimates}C. The estimates of the slip length are obtained from the wall-force variances and the normalized integrated autocorrelations $\tau_F(\tN)$, and our results are reported in Table \ref{table:slip}. The estimates are consistent with those obtained from the conductivity and mobility, demonstrating that we are accurately quantifying the fundamental difference in frictional characteristics between these materials. We note that, when normalized by the static contribution, the two curves very nearly collapse, indicating that the primary contribution to the difference in frictions is due to the wall-force variance, consistent with previous results \cite{Barrat_Bocquet1999_scaling, Falk_et_al2010, Falk_et_al2012, Rajan_hBN_electrostatics, Tocci_et_al2020}. 
There is also an apparent secondary dynamic contribution, revealed by the distinct relaxations of the autocorrelation curves for times less than roughly $1 \, \rm ps$. This dynamic contribution is more clearly illustrated by examining the normalized integrated autocorrelations. The dynamic contribution is then obtained as the difference in the plateau values of these two curves, corresponding to a relaxation time $\tau_F$ that is about $20\%$ smaller in HBN than in GR, compensating some of the static difference.

Several previous studies have connected variations in slip length to variations in contact angle \cite{Huang_Bocquet2008_scaling, Barrat_Bocquet1999_scaling, Barrat_Bocquet1999_PRL, Charlaix_2005_exp}. Huang {\it et al.} \cite{Huang_Bocquet2008_scaling} suggest a quasi-universal scaling relationship relating slip length to contact angle $\theta$ of the form $b \propto \left( 1 + {\rm cos} \theta \right)^{-2}$. This relationship was derived for the case that the difference in solid-liquid interaction strength may be characterized by the difference in magnitude of a single Hamaker constant. This assumes that the magnitude of the in-plane and out-of-plane solid-liquid interaction scale with the same parameter. This is appropriate for spherically symmetric liquid particles interacting with the wall via a central potential, but it is not necessarily the case for polar molecules like water that interact both electrostatically and dispersively, and whose lateral and out-of-plane interactions with the wall may be dominated by different forces.

In order to test this scaling, we must evaluate the wettability of the materials. We probe the out-of-plane, adhesive solid-liquid interactions by examining the reversible work necessary to dewet water from the interface. This calculation is done by placing a thin probe volume $v$ of large cross-sectional area in the vicinity of the interface and calculating the probability $P_v(N)$ of observing $N$ particles in this volume \cite{Patel_et_al2010, Patel_et_al2011_umbrella, Geissler2014_long_wavelength, Godawat_et_al2009}. The reversible work $w_{\rm rev}$ associated with fully evacuating the probe volume is then obtained as $\beta w_{\rm rev} = - {\rm ln} P_v(N = 0)$ \cite{Patel_et_al2010}. In the limit of large $v$, this reversible work reports directly on the liquid-solid surface tension and is thus a direct indication of the effective strength of solid-liquid adhesion.

\begin{figure}[h!]
	\centerline{\includegraphics[scale=0.4]{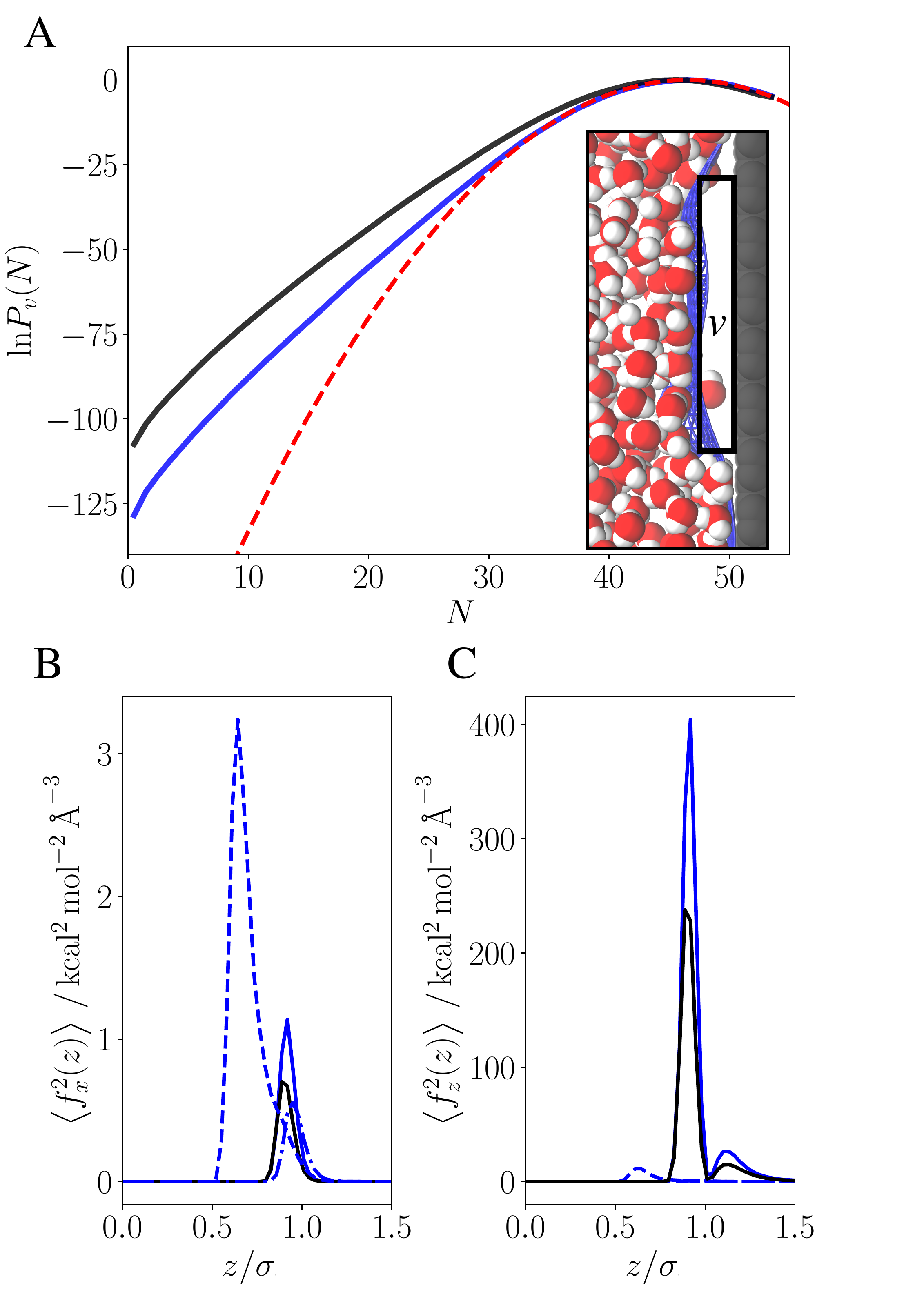}}
    \caption{Lateral and adhesive liquid-solid interaction in HBN (blue) and GR (black). A) The probability $P_v(N)$ of observing $N$ particles in a volume $v$ placed at the solid-liquid interface (inset). The dashed red line indicates a Gaussian fit in the vicinity of the mean particle number. B) The square of the $x$-component of the water-wall force in HBN (blue) and GR (black), including Lennard-Jones contributions (solid lines) and the contributions from the electrostatic interaction of HBN with hydrogen (dashed line) and oxygen (dot-dashed line), as a function of out-of-plane coordinate $z$. C) The corresponding plot for the square of the $z$-components of the water-wall forces. The error bars corresponding to the curves in panels B and C are narrower than the line width and are left out for clarity.}
	\label{fig:contributions}
\end{figure}

We consider a probe volume  $v = 2 \times 2 \times 0.3 \, {\rm nm}^3$ placed at the solid-liquid interfaces in the two channels and compute $P_v(N)$ via the indirect umbrella sampling and the weighted histogram analysis method \cite{Patel_et_al2010, Kumar_et_al1992_WHAM}. The results are indicated in Fig. \ref{fig:contributions}A. We observe that density fluctuations in both materials exhibit Gaussian statistics in the vicinity of the mean particle number. Such fluctuations are characteristic of the bulk fluid \cite{Patel_et_al2011_umbrella}. Far from the mean, the distributions exhibit fat tails, deviating from bulk statistics and from each other. Quantitatively, we find that a change in free energy of $32.1 \, k_\mathrm{B} T/{\rm nm}^2$ in HBN and $26.9 \, k_\mathrm{B} T / {\rm nm}^2$ in GR is needed to dewet the interface, indicating that the free energy of adhesion is stronger in HBN than in GR. If the line tension is negligible \cite{huang_geissler_chandler2001}, the reversible work corresponds to the free energy of interface formation, $w_{\rm rev}/A = \gamma_{\rm LV} ( 1 + {\rm cos} \theta )$, where $\theta$ is the contact angle and $\gamma_{\rm LV}$ is the liquid-vapor surface tension. Assuming the bulk value of $\gamma_{\rm LV} \approx 17.5 \, k_\mathrm{B} T / {\rm nm}^2$, this gives $\theta \approx 30^\circ$ in HBN and $60^\circ$ in GR, qualitatively consistent with recent experimental results indicating contact angles smaller than $90^\circ$ in both materials \cite{hbn_contact_angle, graphene_contact_angle}. We may use these values to evaluate the scaling proposed by Huang {\it et al.} \cite{Huang_Bocquet2008_scaling}, $b \propto (1 + {\rm cos} \theta)^{-2}$. Such a comparison gives an anticipated ratio $b_{\rm GR}/b_{\rm HBN} = \left[\left(1 + {\rm cos} \theta_{\rm HBN}\right)/\left(1 + {\rm cos} \theta_{\rm GR}\right)\right]^2 = \left( w_{\rm rev}^{\rm HBN} / w_{\rm rev}^{\rm GR} \right)^2 \approx 1.6$, far smaller than the ratio of $\sim 6 - 7$ observed here. This suggests that the relatively small difference in wettability of the materials cannot explain the difference in slip lengths. We note that this estimate of the ratio of slip lengths depends only on the ratio of observed reversible works and is thus independent of $\gamma_{\rm LV}$.

In order to understand the difference between the lateral and adhesive water-wall interaction, we examine the profile of the square of the water-wall force $\left\langle {\boldsymbol f}^2(z) \right\rangle \equiv \left\langle \sum_{i = 1}^N {\boldsymbol f}_i^2 \delta \left[ z - z_i (t) \right] \right\rangle$, where ${\boldsymbol f}_i$ is the total force from the wall acting on particle $i$ of the liquid, and ${\boldsymbol f}_i^2 \equiv \left( f_{i,x}^2, f_{i,y}^2, f_{i,z}^2 \right)$ is shorthand for the vector of squared-components of the force. We plot in Figs. \ref{fig:contributions}B and C the $x$- and $z$-components of the squared-force profile, respectively. Whereas the only interaction present between the wall and water in GR is the Lennard-Jones interaction with the oxygen, in HBN we discriminate between three different contributions: the electrostatic forces on hydrogen and oxygen, and the Lennard-Jones force on the oxygen. We see in Fig. \ref{fig:contributions}B that the lateral electrostatic force on the first layer of hydrogens located a distance $\sim \sigma/2 \approx 1.58 \, \rm \AA$, $\sigma$ the Lennard-Jones diameter of oxygen, from the wall greatly exceeds the Lennard-Jones contribution in HBN, as well as the electrostatic force on the first layer of oxygen atoms located at $z \approx \sigma$. This is the layer of hydrogens preferentially oriented towards the nitrogen centers in HBN, as confirmed by the net attraction in the out-of-plane direction experienced by this layer. The typical electrostatic force on hydrogen is nonzero, though much smaller than its peak, up to values of $z \approx \sigma$, indicating a secondary role played by the interaction of the hydrogen atoms oriented such that the O-H bonds in which they participate are parallel to the wall. Figure \ref{fig:contributions}B also indicates that the electrostatic force on the first layer of oxygen, while significantly smaller than that on hydrogen, is comparable to the Lennard-Jones force in both materials. However, our simulations indicate that the frictional response in HBN is set only by the electrostatic interaction of the wall with the first layer of oriented hydrogens. The contribution of the autocorrelation of the Lennard-Jones interaction in HBN to the friction coefficient (Eq. \ref{eqn:lambda}) is almost exactly compensated by the negative cross-correlation of the forces on the hydrogen and oxygen atoms.

The strong confinement of the electrostatic interaction to the first layer of hydrogen atoms (with secondary contributions from the first oxygen layer and the second hydrogen layer) is due to the electroneutrality of the wall, resulting in a strongly spatially confined electric field. The results plotted in Fig. \ref{fig:contributions}B suggest that the enhancement of the friction is due almost entirely to the strong lateral force variance experienced by the layer of hydrogen atoms oriented towards the nitrogen centers. These atoms feel strong local variations in the electrostatic force they experience from the wall owing to their proximity to the wall and the divergent nature of the Coulombic interaction. These variations greatly exceed those induced by the Lennard-Jones corrugation of the wall, and they result in the enhanced friction in HBN.

Figure \ref{fig:contributions}C shows the $z$-component of the squared-force, and it indicates that, in contrast to the lateral case, the Lennard-Jones interactions in the out-of-plane direction dominate the water-wall adhesion, to the point that the out-of-plane electrostatic interaction is entirely negligible. This is again because of the strongly confined nature of the electric field due to the net electroneutrality of the wall, compared to the long-range nature of the attractive portion of the Lennard-Jones force. The dominance of the Lennard-Jones interaction in the solid-liquid adhesion indicates that it is dispersive forces alone that control wettability of both materials in our simulations. This potentially explains the discrepancy recently noted between the pronounced difference in slip characteristics between HBN and GR and the minor differences in wettability \cite{Tocci_et_al2014, Tocci_et_al2020, Rajan_hBN_electrostatics, Wei_and_Luo_2018, Li_and_Zeng_2012, wetting_ref3}. Figure \ref{fig:contributions}C further indicates that the Lennard-Jones interaction is enhanced in HBN, consistent with the smaller contact angle and greater reversible work needed to dewet the interface.

Finally, we evaluate in more detail the relative contributions of static and dynamic effects to the friction coefficients by applying a scaling relationship commonly encountered in the literature \cite{Barrat_Bocquet1999_scaling, Tocci_et_al2014, Falk_et_al2010}. This scaling relationship is obtained by treating the force on the fluid from the fixed walls as an external potential \cite{Barrat_Bocquet1999_scaling}. In this case, we may rewrite $C_{FF}(t)$ in terms of the instantaneous density of species $\alpha$, $\rho_{\alpha}({\bf r},t) = \sum_{i = 1}^{N_{\alpha}} \delta[{\bf r} - {\bf r}_i(t)]$, as
\begin{equation}
C_{FF}(t) = \sum_{\alpha\beta} \int {\rm d} {\bf r} {\rm d} {\bf r}' \, F_{\alpha,x} \left( {\bf r} \right) F_{\beta,x} \left( {\bf r}' \right) \left\langle \rho_{\alpha}({\bf r},t) \rho_{\beta}({\bf r}',0) \right\rangle,
\label{eqn:density_fluct}
\end{equation}
where $F_{\alpha,x} \left( {\bf r} \right)$ is the $x$-component of the force field characterizing the interaction of an isolated atom of species $\alpha$ with the wall, and the sum runs over all the atomic species present in the liquid (hydrogen and oxygen in this case) \cite{Barrat_Bocquet1999_scaling, Tocci_et_al2020}. In general, Eq. \ref{eqn:density_fluct} would contain four terms, proportional to the H-H, O-O and H-O density correlations. However, the only water-wall interaction present in GR is the Lennard-Jones interaction with the oxygen $F_{{\rm LJ}, x} \left( {\bf r} \right)$, reducing this to one term proportional to the O-O density correlation. Furthermore, as noted above, only the electrostatic force on hydrogen $F_{eH, x} \left( {\bf r} \right)$ contributes to the friction in HBN, likewise reducing the expression for $C_{FF}(t)$ in HBN to one term proportional to the H-H density correlation. By decomposing Eq. \ref{eqn:density_fluct} into Fourier components on the reciprocal lattice and inserting the result into Eq. \ref{eqn:lambda}, we obtain
\begin{equation}
\lambda = \beta \int {\rm d} z \, \rho_{\alpha} (z) \sum_{{\boldsymbol k}} \left| F_x \left( {\boldsymbol k} | z \right) \right|^2 \int_0^{\tN} {\rm d} t \, S_{\alpha} \left( k, t | z \right),
\label{eqn:lambda_decomp}
\end{equation}
where neglecting correlations among different $z$ values \cite{Barrat_Bocquet1999_scaling} allows us to rewrite the Fourier decomposition of the density correlation for species $\alpha$ as $A \rho_{\alpha}(z) S_{\alpha}(k, t | z) = \left\langle \sum_{i,j=1}^{N_{\alpha}} {\rm exp} [-i k [x_i(t) - x_j(0)]] \delta[z - z_i(t)] \right\rangle$, where $S_{\alpha}(k, t | z)$ is a two-dimensional time-dependent structure factor conditioned on a value of $z$. We have exploited the isotropy of the contact-layer, observed previously \cite{Falk_et_al2010, Falk_et_al2012} and confirmed in our own simulations, to identify a dependence in the structure factor on the magnitude of $k$ parallel to the wall.

The results plotted in Fig. \ref{fig:contributions}B indicate that the convolution contained in Eq. \ref{eqn:lambda_decomp} is sharply peaked and nonzero only within the contact-layer. We therefore approximate the time-dependent scattering function by its average over the contact-layer and the Fourier amplitude by its value at the peak of the corresponding curve in Fig. \ref{fig:contributions}B. This gives
\begin{equation}
\lambda_{\rm HBN} \approx \frac{\beta}{A} \left\langle N_c^H \right\rangle \left| F_{eH,x}^c \left( {\boldsymbol a} \right) \right|^2 S_{H}^c \left( a \right) \tau_{H}^c \left( a, t_N \right),
\label{eqn:lambda_HBN}
\end{equation}
for HBN, and
\begin{equation}
\lambda_{\rm GR} \approx \frac{\beta}{A} \left\langle N_c^O \right\rangle \left| F_{{\rm LJ},x}^c \left( {\boldsymbol a} \right) \right|^2 S_{O}^c \left( a \right) \tau_{O}^c \left( a, t_N \right).
\label{eqn:lambda_GR}
\end{equation}
where, consistent with previous results, \cite{Falk_et_al2010, Falk_et_al2010} we observe in our simulations that the only relevant component of the Fourier decomposition is that obtained for the shortest vector of the reciprocal lattice, ${\boldsymbol a}$, and truncate the series accordingly. In the above expressions, $\left\langle N_c^{\alpha} \right\rangle$ indicates the average number of atoms of species $\alpha$ in the contact-layer, and $\tau_{\alpha}^c \left( a, \tN \right) \equiv \int_0^{\tN} {\rm d} t \, S_{\alpha}^c \left( a, t \right) / S_{\alpha}^c \left( a, 0 \right)$ indicates the decay time of the density field of species $\alpha$.

The estimates for each of the terms appearing in Eqs. \ref{eqn:lambda_HBN} and \ref{eqn:lambda_GR} are given in Table \ref{table:scaling}. Note that we have divided $\left\langle N_c^H \right\rangle$ by two in recognition of the fact that only one of the two hydrogen atoms in the first layer of waters participates significantly in the electrostatic interaction with the wall. Altogether, this gives for the ratio of slip lengths $b_{\rm GR}/b_{\rm HBN} \approx 6.1$, in very close agreement with the observed ratio. This indicates that the above scaling correctly describes the slip in these systems, and that the friction in HBN is indeed dominated by the electrostatic interaction of the wall with the first, oriented layer of hydrogen atoms. We further learn that the primary contribution to the enhanced friction in HBN is a variance in the lateral force that is an order-of-magnitude larger than that in GR. This is mitigated by reduced structuring of the hydrogen, and by the more rapid relaxation of the hydrogen layer than of the oxygen layer, which are both heavier and experience a Lennard-Jones force that is much weaker than the electrostatic forces experienced by the hydrogen layer \cite{Rajan_hBN_electrostatics}.

\begin{table}
 \begin{center}
  \begin{tabular}{c|c|c|c|c}
	& $\left\langle N_c \right\rangle$ & $|F_x \left( {\boldsymbol a} \right)|^2 \, / ( {\rm kcal}/ {\rm mol} \,{\rm \AA})^{-2}$ & $S \left( a \right)$ & $\tau \left( a , \tN \right) \, / \, {\rm ps}$ \\ \hline
	HBN & $147.8$ & $0.928$ & $0.880$ & $0.338$ \\
	GR & $155.4$ & $0.0981$ & $1.215$ & $0.364$	 \\
  \end{tabular}
  \caption{Estimates of terms contributing to scaling of friction coefficients in HBN and GR.}
  \label{table:scaling}
 \end{center}
\end{table}

Our results indicate the crucial importance that material chemistry is expected to exert on flow properties in nanoconfinement, and they suggest that relatively moderate changes in chemistry can have profound implications for the slip length even within crystallographically similar materials. In this study, we have investigated atomically smooth channels and illustrated the particular role of Coulombic interactions in increasing the scale of lateral interactions between water and HBN over that of water and GR without significantly modifying the wettability. These findings suggest that surface charge adsorption on chemically activated crystals should likewise have a dramatic effect on fluid slip. This influence is in addition to the key role surface charge plays in the generation of boundary-driven flow and conversion of osmotic energy into useful electrical or mechanical work \cite{Siria_et_al2017,coming_of_age}. This is particularly relevant in light of recent results highlighting the difference in charge adsorption mechanisms in GR and HBN and the large surface charges obtainable in activated HBN \cite{Siria2013_BN,Grosjean_BN}. Accordingly, investigating the interplay between defects, surface charge, and fluid slip is a key next step in the design of ideal materials for blue energy generation.
 
\vspace{0.5cm}
\noindent{\bf Acknowledgments}
A. R. P and D. T. L. were supported by the LDRD program at LBNL under U. S. Department of Energy Office of Science, Office of Basic Energy Sciences under Contract No. DE-AC02-05CH11231.
\vspace{0.5cm}

\noindent{\bf References}
\bibliography{slip_ref}

\end{document}